\documentclass[a4paper]{jpconf}
\usepackage{graphicx}

\begin{document}
\title{The linearity of quantum mechanics from the perspective of Hamiltonian cellular 
automata}

\author{Hans-Thomas Elze}

\address{Dipartimento di Fisica ``Enrico Fermi'', Universit\`a di Pisa,  
Largo Pontecorvo 3, I-56127 Pisa, Italia}

\ead{elze@df.unipi.it}

\begin{abstract}
We discuss the action principle and resulting Hamiltonian equations of motion 
for a class of integer-valued cellular automata  introduced recently 
\cite{PRA2014}. Employing sampling theory, these  deterministic finite-difference
equations 
are mapped reversibly on continuum equations describing a set of bandwidth limited 
harmonic oscillators. They  represent the Schr\"odinger equation.  However, 
modifications reflecting  the bandwidth limit are incorporated, {\it i.e.}, the presence of a   
time (or length)  scale. When this discreteness scale is taken to zero, the usual results are 
obtained. 
Thus, the linearity of quantum mechanics can be traced to the postulated action principle 
of such cellular automata and its conservation laws to discrete ones. The cellular automaton conservation laws are in one-to-one correspondence with those of the 
related quantum mechanical model, while admissible symmetries are not.
\end{abstract}

\section{Introduction} 
The linearity of quantum mechanics (QM) is a fundamental feature most notably 
embodied in the Schr\"odinger equation. This linearity does not depend on the 
particular object  under study, provided it is sufficiently isolated from 
anything else. It is naturally reflected in the superposition principle and   
entails the ``quantum essentials'' interference and entanglement, without which 
modern applications of QM,  
{\it e.g.} in precision measurement and information technologies, 
and the ongoing study of the foundations of QM would not be the same.    

The linearity of QM has been questioned now and then and various nonlinear modifications have been proposed, in order to test experimentally the  
robustness of QM against such nonlinear deformations. 
This has been thoroughly discussed by Jordan presenting a stepwise proof that QM 
has to be linear based on the separability  
assumption {\it ``... that the dynamics we are considering can be independent
of something else in the universe, that the system we are considering can be described as
part of a larger system without interaction with the rest of the larger 
system''}\,\cite{Jordan}. 
It is also worth recalling that most proposals of nonlinearities have 
been beset with the problem of superluminal 
signalling or communication between branches of the wave function, see, {\it e.g.},    
Refs.\,\cite{Weinberg,Gisin, Polchinski,me08}; 
since then, ``no signalling'' has become an important criterium that attempted modifications of QM are confronted with. 

We have recently provided a different point of view concerning the linearity of 
QM \cite{PRA2014}.  We have related QM to the mechanics of a class of Hamiltonian cellular automata. Which shows that the linearity of QM is necessarily required  
by the consistency of the postulated action principle for such discrete dynamics. 
This may further  the  understanding of interference, entanglement, and    
measurement in QM and lead to new approximation schemes in quantum theory.  

Our approach was motivated by explorations of discrete deterministic mechanics 
by Lee \cite{Lee}, by the study of bandwidth limited fields and their possible role in   
the physics of gravity and spacetime by Kempf \cite{Kempf},  and by the representation 
of QM in terms of classical notions (observables, phase space, Poisson bracket algebra)   
by Heslot \cite{Heslot85,me12}. Presently, we recover our earlier arguments, which  
combine these ideas, and add a few details on the way. 

\section{Discrete Hamiltonian mechanics}
Discreteness has many facets in physics, besides quantization, {\it e.g.},  
discrete maps for numerical studies of complex systems, regularized versions of quantum field theories on spacetime lattices, or intrinsically discrete processes. 
Finite difference equations are expected to play a prominent role here, instead of the 
usual preponderance of differential calculus.   

Lee and collaborators proposed to incorporate  fundamental discreteness into all of dynamics \cite{Lee} (and references therein). This was apparently motivated  
by the difficulties encountered in trying to formulate a consistent  
theory of ``quantum gravity'', or even ``the'' unified theory.   
Thus, deterministic discrete mechanics 
derives from the assumption that {\it time is a discrete dynamical variable}. This 
invokes a {\it fundamental time or length scale} (in natural units), $l$, and can be 
rephrased that in a fixed $(d+1)$-dimensional spacetime volume $\Omega$ maximally $\Omega /l^{d+1}$ measurements can be performed or this number of events take place 
\cite{Lee}. 

We consider the new discreteness scale $l$  in the spirit of {\it deformations} of Lorentz symmetry, in the form of ``doubly special relativity'' (DSR) \cite{Hossenfelder}, 
or in its explicit breaking, and in the nonlinear {\it deformation} of QM 
\cite{Weinberg,Gisin,Polchinski,me08} mentioned earlier. While such studies 
necessarily introduce additional parameters, the aim is to probe the stability of 
the existing theories, the Standard Model and QM in particular, against such 
deformations, predicting new phenomena that eventually could 
lead  to a deeper theory with a smaller set of fundamental parameters 
\cite{DuffOkunVeneziano}. It is commonly  
expected that $l\equiv l_{Pl}$, {\it i.e.}, that discreteness and Planck scale coincide. 
 
Various discrete models have been elaborated, which share desirable symmetries 
with the corresponding continuum theories  while presenting finite degrees of freedom. 
Different forms and (dis)advantages 
of a Lagrangian formulation  \cite{Lee} have been discussed, 
{\it e.g.}, in 
Refs.\,\cite{DInnocenzo87,HongBin05,Toffoli11}. In the following, we introduce an  action  
principle instead which leads to particularly transparent and symmetric Hamiltonian equations of motion, corresponding to a discrete phase space picture. 

\section{The action principle for Hamiltonian cellular automata}
The state of a classical cellular automaton (CA) with a denumerable set of degrees 
of freedom will be represented by {\it integer-valued} ``coordinates''   
$x_n^\alpha ,\tau_n$ and ``conjugated momenta'' $p_n^\alpha ,\pi_n$, where 
$\alpha\in {\mathbf N_0}$ denote different degrees of freedom and $n\in {\mathbf Z}$  
different states. 

The {$x_n$ and $p_n$ might be higher dimensional vectors, while $\tau_n$ and 
${\cal P}_n$ are assumed one-dimensional. In separating the ``coordinate'' $\tau_n$ 
from the $x_n^\alpha$'s (and correspondingly $\pi_n$ from the $p_n^\alpha$'s), 
we follow Refs.\,\cite{Lee,DInnocenzo87,HongBin05}. This 
degree of freedom represents the {\it dynamical time} variable here,  
instead of  the external parameter time of Newtonian mechanics or QM. 
   
Finite differences, for all dynamical variables, are defined by 
\begin{equation}\label{findiff}
\Delta f_n:=f_n-f_{n-1} 
\;\;. \end{equation} 
Furthermore, we define (using henceforth the summation convention for Greek indices, 
$r^\alpha s^\alpha\equiv\sum_\alpha r^\alpha s^\alpha$): 
\begin{eqnarray}\label{CAHamiltonian} 
{\cal A}_n&:=&\Delta \tau_n (H_n+H_{n-1})+a_n 
\;\;, \\ [1ex] \label{CAHamiltonian1}
H_n&:=&\frac{1}{2}S_{\alpha\beta}(p_n^\alpha p_n^\beta+x_n^\alpha x_n^\beta  )
+A_{\alpha\beta}p_n^\alpha x_n^\beta + R_n 
\;\;, \\ [1ex] \label{CAHamiltonian2}
a_n&:=&c_n\pi_n
\;\;, \end{eqnarray}  
where constants, $c_n$, and symmetric, $\hat S\equiv\{ S_{\alpha\beta}\}$,  and antisymmetric, $\hat A\equiv\{ A_{\alpha\beta}\}$, matrices are all integer-valued;  
$R_n$ stands for higher than second powers in $x_n^\alpha$ or  $p_n^\alpha$.  
The choice of the right-hand side of Eq.\,(\ref{CAHamiltonian2}) determines the 
behaviour of the variable $\tau_n$; for our present purposes, a very simple 
choice suffices and will be further discussed shortly (nontrivial 
`potential' terms have been made use of in Ref.\,\cite{me13}). 
 
In terms of the definitions just given, we introduce  the integer-valued CA {\it action}  
\begin{equation}\label{action} 
{\cal S}:=
\sum_n[(p_n^\alpha +p_{n-1}^\alpha )\Delta x_n^\alpha 
+(\pi_n+\pi_{n-1})\Delta\tau_n
-{\cal A}_n]  
\;\;, \end{equation} 
and postulate that a Hamiltonian CA is described by the following  
\vskip 0.25cm 
{\it Action Principle:} \hskip 0.1cm 
The CA obeys the discrete updating rules 
(equations of motion)   which  are determined by $\delta {\cal S}=0$,  
referring to arbitrary integer-valued variations of all dynamical variables,  
\begin{equation}\label{variation} 
\delta g(f_n):=[g(f_n+\delta f_n)-g(f_n-\delta f_n)]/2 
\;\; , \end{equation} 
where $f_n$ stands for one of the variables on which  polynomial $g$ may depend.\,$\bullet$ \vskip 0.25cm
 
Several remarks are in order here. -- We observe that the variations  of 
constant, linear, or quadratic terms  yield results that are analogous to the continuum 
case. -- While infinitesimal variations do not conform with integer valuedness, 
there is no {\it a priori}  constraint on integer ones. However, for arbitrary  
$\delta f_n$, the {\it remainder of higher powers} in 
Eq.\,(\ref{CAHamiltonian1}), which enters the action, has to vanish for consistency, 
$R_n\equiv 0$. Otherwise the number of equations of motion 
generated by the action principle, generally, would exceed the number of 
variables. We note that a suitably chosen $R_0$ or a 
sufficient small number of such remainder terms could encode the 
necessary {\it initial conditions} for the CA evolution.      
 
With the notation $\dot O_n:=O_{n+1}-O_{n-1}$,  
the following CA {\it equations of motion} are obtained:   
\begin{eqnarray}\label{xdotCA} 
\dot x_n^\alpha &=&\dot\tau_n(S_{\alpha\beta}p_n^\beta +A_{\alpha\beta}
x_n^\beta ) 
\;\;, \\ [1ex] \label{pdotCA} 
\dot p_n^\alpha &=&-\dot\tau_n(S_{\alpha\beta}x_n^\beta -A_{\alpha\beta}p_n^\beta ) 
\;\;, \\ [1ex] \label{taudotCA} 
\dot\tau_n&=&c_n 
\;\;, \\ [1ex] \label{pidotCA} 
\dot\pi_n&=&\dot H_n 
\;\;, \end{eqnarray} 
which are discrete analogues of Hamilton's equations, where all terms are 
integer-valued. Discreteness of the  {\it automaton time} $n$ is reflected in the    
finite difference equations here.  

Note that  the $\dot\tau_n$ 
present {\it background} parameters for the evolving $x,p$-variables, as  
a consequence of Eqs.\,(\ref{CAHamiltonian2}), (\ref{taudotCA}). 
Generally, $\dot\tau$ is a    
{\it lapse function} in Eqs.\,(\ref{xdotCA})--(\ref{pdotCA}).  

The Eqs.\,(\ref{xdotCA})--(\ref{pidotCA}) are {\it time reversal invariant}; 
the state $n+1$ can be calculated from knowledge of the 
earlier states $n$ and $n-1$ and the state $n-1$ from the later ones $n+1$ and $n$. 

\section{Conservation laws and CA symmetries} 
Surprisingly, there are conservation laws that are 
always respected by the discrete equations of motion Eqs.\,(\ref{xdotCA}) and (\ref{pdotCA}). -- Introducing the   
self-adjoint matrix $\hat H:=\hat S+i\hat A$,  
these equations can be combined into: 
\begin{equation}\label{discrS} 
 \dot x_n^\alpha +i\dot p_n^\alpha =-i\dot\tau_n H_{\alpha\beta}
(x_n^\beta +ip_n^\beta ) 
\;\;, \end{equation} 
and its adjoint. Thus, we recover a {\it discrete analogue of Schr\"odinger's equation}, 
with $\psi_n^\alpha :=x_n^\alpha +ip_n^\alpha$ as the amplitude of the 
``$\alpha$-component''  
of ``state vector'' $|\psi\rangle$ at ``time'' $n$. Then, the 
Eqs.\,(\ref{xdotCA})--(\ref{pdotCA}) imply this:  
\vskip 0.25cm 
{\it Theorem A:} \hskip 0.1cm For any matrix $\hat G$ that commutes with $\hat H$, 
$[\hat G,\hat H]=0$, there 
is a {\it discrete conservation law}: 
\begin{equation}\label{Gconserv} 
 \psi_n^{\ast\alpha}G_{\alpha\beta}\dot\psi_n^\beta +
\dot\psi_n^{\ast\alpha}G_{\alpha\beta}\psi_n^\beta =0 
\;\;. \end{equation}  
For self-adjoint $\hat G$,  with complex integer elements, 
this relation concerns real integer quantities.\,$\bullet$  
\vskip 0.25cm 
{\it Corollary A:} \hskip 0.1cm For $\hat G:=\hat 1$, the Eq.\,(\ref{Gconserv}) implies 
a {\it conserved constraint} on the state variables:  
\begin{equation}\label{psiconserv} 
 \psi_n^{\ast\alpha}\dot\psi_n^\alpha +
\dot\psi_n^{\ast\alpha}\psi_n^\alpha =0 
\;\;. \end{equation}  
For $\hat G:=\hat H$, an {\it energy conservation} law follows.\,$\bullet$  
\vskip 0.25cm 

Note that Eqs.\,(\ref{Gconserv}) and (\ref{psiconserv}) {\it cannot} be trivially 
``integrated'', since the {\it Leibniz rule} is modified. Recalling  
$\dot O_n:=O_{n+1}-O_{n-1}$, we have, for example, 
$O_{n+1}O'_{n+1}-O_{n-1}O'_{n-1}=\frac{1}{2}(\dot  O_n[O'_{n+1}+O'_{n-1}]
+[O_{n+1}+O_{n-1}]\dot O'_n)$, instead of the product rule of differentiation. 

Furthermore, we cannot  obtain a continuum limit simply by letting  
the discreteness scale $l\rightarrow 0$, 
as for example in   Refs.\,\cite{Lee,me13}.   
integer valuedness here conflicts with continuous time 
and related derivatives. 

\subsection{Another CA action and conservation laws without admissible unitary symmetries?}  
The CA action is invariant under suitable unitary transformations. This  
can be most easily 
recognized by considering an equivalent form of the action, {\it i.e.}, which 
generates the same discrete equations of motion as before. 
We may replace the definition given in Eq.\,(\ref{action}) by: 
\begin{equation}\label{actionPsi} 
{\cal S}:=
\sum_n[\mbox{Im}(\psi_n^{\alpha\;*}\psi_{n-1}^\alpha )  
+(\pi_n+\pi_{n-1})\Delta\tau_n
-{\cal A}_n]  
\;\;, \end{equation}  
with $\psi_n^\alpha :=x_n^\alpha +ip_n^\alpha$ and Im$X:=(X-X^*)/2i$, 
together with the replacement of the 
definition of $H_n$ (recall $R_n\equiv 0$), Eq.\,(\ref{CAHamiltonian1}), by: 
\begin{equation}\label{HnPsi} 
H_n:=\frac{1}{2}H_{\alpha\beta}\psi_n^{\alpha\;*}\psi_n^\beta 
\;\;, \end{equation} 
which enters the action through ${\cal A}_n$, Eq.\,(\ref{CAHamiltonian}). Then, 
$n$-independent unitary transformations $\hat U$, with $\psi_n'=\hat U\psi_n$ and 
$[\hat U,\hat H]=0$, leave the action ${\cal S}$ invariant. 

The self-adjoint matrices $\hat G$ of {\it Theorem A} generate     
unitary transformations which leave ${\cal S}$ invariant. 
However, since the CA variables $\psi_n^\alpha :=x_n^\alpha +ip_n^\alpha$ 
are restricted to be {\it complex integer-valued}, only unitary transformations 
that preserve this property are {\it admissible}. In general, we expect that there are 
very few interesting ones. Thus, we encounter here the situation that there can 
exist CA conservation laws, according to {\it Theorem A} or {\it Corollary A}, 
which are {\it not} 
related to {\it admissible symmetry transformations}. This differs 
from from what is usually the case in QM.

\section{Sampling theory} 
It is worth recalling the underlying assumption of discrete mechanics 
that the density 
of events and, thus, of information content of spacetime regions 
is cut off by the scale $l$ \cite{Lee,SorkinDICE02}.  
Therefore, despite the observed similarities of our Hamiltonian CA with QM systems, 
we may wonder whether the discreteness of a deterministic CA can be reconciled with 
any continuum description at all and, in particular, with QM?  

Indeed, we have argued that physical fields, wave functions in particular, could be 
{\it simultaneously discrete and continuous}, represented by sufficiently smooth  functions containing a finite density of degrees of freedom \cite{PRA2014}. 
This idea has recently been introduced by Kempf 
and has led to a covariant ultraviolet cut-off suitable for theories 
including gravity \cite{Kempf}.  
However, neither {\it integer-valued CA} nor the {\it structure of QM} have been 
addressed in this way. 

The fact that information can have simultaneously
continuous and discrete character has been pointed out by Shannon 
in his pioneering work \cite{Shannon}. This is routinely applied in signal 
processing, converting analog to  
digital encoding and {\it vice versa}. Sampling theory demonstrates  
that a bandlimited signal can be perfectly reconstructed, 
provided discrete samples of it are taken at the rate of at least twice the band 
limit (Nyquist rate). For an extensive review and modern developments of the 
theory, see Refs.\,\cite{Jerri,StrohmerTanner}, respectively. 

We consider the {\it Sampling Theorem} in its 
simplest form \cite{Kempf,Jerri}:  
Consider square integrable {\it bandlimited functions} $f$, {\it i.e.}, which can be 
represented as $f(t)=(2\pi )^{-1}\int_{-\omega_{max}}^{\omega_{max}}\mbox{d}\omega\; 
\mbox{e}^{-i\omega t}\tilde f(\omega )$, with bandwidth $\omega_ {max}$. Given 
the set of amplitudes $\{ f(t_n)\}$ for the set  $\{ t_n\}$ of equidistantly spaced times  
(spacing $\pi /\omega_{max}$), the function $f$ is obtained for all $t$ 
by: 
\begin{equation}\label{samplingtheorem} 
f(t)=\sum_n f(t_n)\frac{\sin [\omega_{max}(t-t_n)]}{\omega_{max}(t-t_n)} 
\;\;. \end{equation} 

Since the CA time is given by the integer $n$, the corresponding discrete {\it physical time}  
is obtained by  multiplying with the fundamental scale $l$,  $t_n\equiv nl$, and the 
bandwidth by $\omega_{max}=\pi /l$. 

When attempting to {\it map invertibly} Eqs.\,(\ref{xdotCA})--(\ref{pdotCA}) on reconstructed 
continuum equations, according to Eq.\,(\ref{samplingtheorem}), the nonlinearity 
on the right-hand sides is problematic: the product of two 
functions, with bandwidth $\omega_ {max}$ each, is not a function with the same 
bandwidth. Therefore, we presently assume that $\dot\tau_n$ is a 
constant and postpone consideration of more general situations.  

Let us recall Eq.\,(\ref{discrS}). Inserting $\psi_n^\alpha :=x_n^\alpha +ip_n^\alpha$ and      
applying the {\it Sampling Theorem}, this discrete time equation
is mapped to the  {\it continuous time equation}: 
\begin{equation}\label{modS}
\frac{\hat D_l-\hat D_{-l}}{2}\psi^\alpha (t)=\sinh (l\partial_t)\psi^\alpha (t)
=\frac{1}{i}H_{\alpha\beta}\psi^\beta (t) 
\;, \end{equation} 
where we employed the translation operator defined by $\hat D_Tf(t):=f(t+T)$ and 
set $\dot\tau_n\equiv\dot\tau =2$. 

Thus, we obtain the {\it Schr\"odinger equation}, however, modified in important ways. 
(We use the QM terminology freely and concentrate on new effects arising here.)  
The wave function $\psi^\alpha$ has bandwidth $\omega_{max}$, due to 
reconstruction formula (\ref{samplingtheorem}). This leads to an 
{\it ultraviolet cut-off} of the energy $E$ of stationary states of the generic form $\psi_E(t):=\exp (-iEt)\tilde\psi$. Diagonalizing the self-adjoint  Hamiltonian,  
$\hat H\rightarrow\mbox{diag}(\epsilon_0,\epsilon_1,\dots )$,  
Eq.\,(\ref{modS}) yields the eigenvalue equation:
\begin{equation}\label{eigenS} 
\sin (E_\alpha l)=\epsilon_\alpha  
\;\;, \end{equation} 
and a {\it modified dispersion relation}, 
$E_\alpha =l^{-1}\arcsin (\epsilon_\alpha )=
l^{-1}\epsilon_\alpha [1+\epsilon_\alpha^{\;2}/3!+\mbox{O}(\epsilon_\alpha^{\;4})]$ 
\cite{disprel}.   
The spectrum $\{ E_\alpha\}$ is cut off by 
the condition $|\epsilon_\alpha |\leq 1$, entailing   
$|E_\alpha |\leq \pi /2l=\omega_{max}/2$, {\it i.e.} half the bandlimit.  

For CA with a finite (or truncated) number of degrees of freedom or states, labeled 
by $\alpha$, the constant $\dot\tau$ could be determined instead by the largest  
eigenvalue of $H$, 
$\epsilon_{max}:=\mbox{max}\{ |\epsilon_\alpha |\}$, choosing 
$\dot\tau =1/\epsilon_{max}$. In this way, there would be stationary states 
corresponding to all eigenvectors of $\hat H$. Otherwise, states corresponding to 
larger eigenvalues  of $\hat H$ are unstable, related to complex solutions of the 
dispersion relation.   

The modified Schr\"odinger equation (\ref{modS}) incorporates an infinite 
series of higher-order time derivatives. These are negligible for low-energy wave functions, which vary little with respect to the cut-off scale, {\it i.e.}    
$|\partial^k\psi /\partial t^k|\ll l^{-k}=(\omega_{max}/\pi )^k$. However, 
in general, they would require an infinity of initial data.  

\subsection{Continuous time conservation laws}  
The relation between Eq.\,(\ref{discrS}) and Eq.\,(\ref{modS}), together 
with the linearity of both equations, suggest that the correct continuous time 
conservation laws can be obtained by the replacement
\begin{equation}\label{cconserv}   
\dot\psi_n:=\psi_{n+1}-\psi_{n-1}\;\;\longrightarrow\;\;\frac{1}{i}\sin (il\partial_t)\psi (t)
\;\;, \end{equation} 
from Eqs.\,(\ref{Gconserv}) and (\ref{psiconserv}), respectively. Indeed, by Eq.\,(\ref{modS}),  
the following holds: \vskip 0.25cm 
{\it Theorem B:} \hskip 0.1cm For any matrix $\hat G$ with $[\hat G,\hat H]=0$, 
there is a {\it continuous time conservation law}: 
\begin{equation}\label{cGconserv} 
 \psi ^{\ast\alpha}G_{\alpha\beta}\sin (il\partial_t)\psi^\beta +
[\sin (il\partial_t)\psi^{\ast\alpha}]G_{\alpha\beta}\psi^\beta =0 
\;\;, \end{equation}  
in particular,     
\begin{equation}\label{cpsiconserv} 
 \psi ^{\ast\alpha}\sin (il\partial_t)\psi^\alpha +
[\sin (il\partial_t)\psi^{\ast\alpha}]\psi^\alpha =0 
\;\;, \end{equation}  
which modifies the QM   
wave function {\it normalization}, referring to a basis labeled by $\alpha$.\,$\bullet$ \vskip 0.25cm 

Only now we can remove the ultraviolet cut-off,  with $l\rightarrow 0$, and recover 
familiar QM results from the leading order terms in 
Eqs.\,(\ref{cGconserv})--(\ref{cpsiconserv}). (If $l$ is a fundamental {\it constant}, 
this limit may be interesting for heuristic reasons alone.)   

For example, consider the real symmetric {\it two-time function}, 
\begin{equation}\label{C}
2C_{\hat G}(t_1,t_2):=\psi ^{\ast\alpha}(t_1)G_{\alpha\beta}\psi^\beta (t_2)\; +\;
\mbox{c.c.}
\;\;, \end{equation}
where $X+\mbox{c.c.}:=X+X^\ast$ and 
$\hat G$ is a self-adjoint matrix, with $[\hat G,\hat H]=0$. Applying {\it Theorem B}, 
we obtain: \vskip 0.25cm  
{\it Corollary B:} \hskip 0.1cm The two-time function $C_{\hat G}$ is invariant under 
discrete translations of this form: 
\begin{equation}\label{Ctransl} 
C_{\hat G}(t-l,t)=C_{\hat G}(t,t+l) 
\;\;, \end{equation} 
implying that it is fixed everywhere by giving  $C_{\hat G}(t,t+l)$ for all $t$ in an interval
$[t_0,t_0+l[$. \,$\bullet$ \vskip 0.25cm  
The wave function normalization,  
$\psi^{\ast\alpha}\psi^\alpha =1$,  
then arises here from the coincidence limit of a   
two-time function with the property $C_{\hat 1}(t,t+l)\equiv 1$, for all $t$: 
\begin{equation}\label{wfnorm} 
1=\lim_{l\rightarrow  0}C_{\hat 1}(t,t+l)=\psi^{\ast\alpha}(t)\psi^\alpha (t) 
\;\;, \end{equation} 
which is consistent with Eq.\,(\ref{cpsiconserv}) and  essential for the  probability interpretation in QM. An analogous {\it equal-time} constraint, in general,      
does not exist on the CA level of description. {\it E.g.},  
$\psi_n^{\ast\alpha}\psi_n^\alpha =x_n^\alpha x_n^\alpha +p_n^\alpha p_n^\alpha =1$, instead of Eq.\,(\ref{psiconserv}), is compatible only with rather trivial evolution, since 
all variables are integer-valued.  
 
It is remarkable how properties of Hamiltonian CA produce familiar QM results, 
even if modified by the finite scale $l$.  
The operators or matrices that generate the QM conservation laws  
do so for the bandwidth limited continuum theory as well, 
as stated by {\it Theorem B}. Since the {\it same} vanishing commutator is responsible for the CA conservation laws, Eqs.\,(\ref{Gconserv})--(\ref{psiconserv}), they correspond to each other one-to-one. Yet the 
QM symmetry transformations, generally, comprise a larger 
set than the admissible discrete ones for CA, which have to respect complex integer 
valuedness of the dynamical variables. 

These observations leave us with an intriguing question: {\it What, if any, would be 
physical reasons for the existence of Hamiltonian CA conservation laws that are not 
tied to symmetries, which are fully developed only in the continuum limit?} 

\section{Discussion}
It will be important to extend the CA--QM map to  
relativistic QM and QFT. Since wave equations and functional 
Schr\"odinger equation are linear and have a Hamiltonian formulation, it should 
be possible to employ a generalized {\it Sampling Theorem} for fields.  
It has been shown indeed that the d'Alembert operator can be covariantly 
regularized by imposing a finite 
bandwidth of its spectrum \cite{Kempf}, a useful ingredient. -- 
Conversely, given a {\it Hamiltonian} CA, one could invoke the path integral 
for classical  systems  \cite{us} plus reconstruction formulae, attempting to obtain  
a relativistic bandwidth limited quantum (field) theory.   
-- Other constructions of CA for  
{\it relativistic models} have appeared, which either incorporate QM features from the  outset, {\it e.g.}, for the Dirac equation 
\cite{FeynmanIBBdArianoArrighi}, or derive them, {\it e.g.}, for bosonic QFT and 
a superstring model \cite{tHooft}; see also references there. 
All these models have been {\it noninteracting}. 
 
Lack of interactions there might be dictated by additional 
restrictions, such as locality, when placing a CA, say at Planck scale, 
into physical spacetime as experienced at the scales   
where QM is tested. It is remarkable that arbitrary QM $N$-level systems 
can be described by $2N-1$ nonrelativistic coupled 
oscillators in one fictitious space dimension  \cite{Skinner13}. Could it be   
that fundamental CA exist in an abstract space and  
QM and spacetime emerge together from there? 
 
The nonrelativistic Hamiltonian CA {\it do incorporate 
interactions} through  the matrix elements $H_{\alpha\beta}$.  
Their $x^\alpha,p^\alpha$- variables can be embedded 
into two-dimensional phase space, similarly as in Ref.\,\cite{Skinner13}. Yet  
other interpretations are possible, such as $\alpha$ labelling sites of a $d$-dimensional 
lattice or elements of the Hilbert space arrived at in   
the QM description. 
This freedom appears in the nonrelativistic situation studied here without  
reference to gravitation or dynamical spacetime.  

Interactions are also incorporated in a statistical theory of certain matrix models, 
which leads to QM behaviour emerging from a Gibbs distribution \cite{Adler}. 
Similarly as in Refs.\,\cite{tHooft}, this assumes a very 
particular form of dynamics  and it remains to 
be seen whether gauge theories as, for example,  in the Standard Model can 
be recovered. -- In distinction, we do not make any assumptions 
about particular interactions but explore the mapping between structural 
features of Hamiltonian CA and of QM. It will be challenging to identify   
principles that distinguish a physically relevant Hamiltonian and related 
conservation laws within an ``ontology'' of CA.    

In Ref.\,\cite{PRA2014}, we have discussed how the essential features of {\it entanglement}  in QM and {\it apparent nonlocality}  \cite{Englert} come into play in our 
approach. It will be most interesting to reconsider questions of entanglement and 
locality in a relativistic generalization of the present theory. -- 
Observables, measurements, and Born rule can be discussed in the bandwidth limited 
theory with help of Heslot's work \cite{Heslot85} and implications 
for CA {\it per se} deserve further study.   

We remark that Hamiltonian CA might be useful to simulate complex QM 
systems by the mapping on computer friendly {\it integer-valued}  dynamical 
variables. A few hints in this direction have been mentioned in Ref.\,\cite{PRA2014}.  

Let us also draw attention to the differences between the presently introduced 
Hamiltonian CA and quantum cellular or {\it quantum lattice-gas automata} 
(QLGA). The QLGA have recently attracted attention, since 
they are, by construction, discretizations of the Schr\"odinger equation 
\cite{Meyer97,Boghosian98}, cf. also \cite{FeynmanIBBdArianoArrighi}. 
They are specifically made to reproduce a quadratic kinetic 
energy term in the Schr\"odinger equation in the continuum limit in configuration 
space. This involves judicious choice of 
transformation matrices, i.e. implicitly of a number of dimensionless parameters 
\cite{Boghosian98}. 
However, the Hilbert space structure of QM state space 
with complex wave functions and linearity and unitarity of their evolution have to  
be incorporated  {\it ab initio}. 

In these respects, our approach differs remarkably: It is 
based on {integer-valued} dynamical variables and an underlying {action principle}. 
This  {\it implies} linearity and unitarity together with all conservation laws. 
Unlike the case of QLGA and in accordance with the discussion in Sect.\,II. of  
the role of the discreteness scale $l$, the Hamiltonian CA here 
provide a {\it discrete deformation} of QM that reduces to it for $l\rightarrow 0$. 

\section{Conclusion}
A map between Hamiltonian cellular automata (CA) and quantum mechanics (QM) 
has been constructed by combining  
elements of {\it discrete mechanics} \cite{Lee}, {\it sampling theory} \cite{Kempf}, 
and {\it Hamiltonian formulation}  of QM \cite{Heslot85}. Thus, structural 
features of QM, the Schr\"odinger picture of evolution and conservation laws in 
particular, can be seen to originate in  
integer-valued CA incorporating a fundamental scale. 

The postulated action 
principle refers to arbitrary integer variations of the dynamical CA variables.  
This enforces the {\it linearity} of the theory, as we have demonstrated  \cite{PRA2014}. 

Consequently, the separability 
assumption mentioned in Sect.\,I., which underlies the derivation 
of the linearity of QM {\it within} its formal framework, 
can here be substituted as follows: \\ 
{\it ``... the dynamics we are considering can be independent of something else in the 
universe''} \cite{Jordan}, if and only if {\it the} CA  
{\it action is stationary under arbitrary integer variations}.   

This provides a new view of linearity and the superposition principle in 
quantum mechanics with manifold consequences and generalizations to be explored. 

\ack

I wish to thank G.\ 't Hooft for discussions and correspondence and G.\ Gr\"ossing and his collaborators and J.\ Walleczek for organizing the symposium EmQM13 ``Emergent Quantum Mechanics'' and for the kind hospitality in Vienna supported by Fetzer Franklin Fund. 


\end{document}